\documentclass[12pt]{iopart}
\usepackage{epsf}

\begin{document}

%\title[]{Analytic treatment of non-Gaussianity in multi-field stochastic inflation}
\title[]{Non-Gaussianity in Multi-field Stochastic Inflation \\
with the Scaling Approximation}

%\author{Neil Scriven\dag\ and Romneya Robertson\ddag  
%\footnote[3]{To
%whom correspondence should be addressed (romneya.robertson@iop.org)}}

\author{Takeshi Hattori and Kazuhiro Yamamoto}
\address{
Graduate School of Sciences, Hiroshima University, 
Higashi-hiroshima, 739-8526, Japan}

%\address{\dag\ Production Editor, Institute of Physics 
%Publishing, Dirac
%House, Temple Back, Bristol BS1 6BE, UK}
%\address{\ddag\ Electronic Services Specialist, 
%Institute of Physics Publishing, 
%Dirac House, Temple Back, Bristol BS1 6BE, UK}

\begin{abstract}
The statistics of multi-field inflation are investigated using 
the stochastic approach. 
We analytically obtain the probability distribution function 
of fields with the scaling approximation by extending the 
previous work by Amendola. The non-Gaussian nature of the 
probability distribution function is investigated decomposing the 
fields into the adiabatic and isocurvature components. 
We find that the non-Gaussianity of the isocurvature 
component can be large compared with that of the 
adiabatic component. The adiabatic and isocurvature components
may be correlated at nonlinear order in the skewness and kurtosis
even if uncorrelated at linear level.
\end{abstract}

%\maketitle
%-----------------------------------------------------------
\section{Introduction}
%-----------------------------------------------------------
The inflationary universe is now the standard paradigm that
explains the observed universe naturally \cite{Guth,Sato}. 
One of the most important predictions of inflation is the 
primordial fluctuations from which the large scale 
structure of the universe originates. 
Many models of the inflation have been proposed.
Multi-field inflation is a natural alternative model 
to the standard single field inflaton model (e.g., \cite{MS,ST,Gordon,AGWS,ATP}). 
In such the multi-field inflation model, the isocurvature 
perturbation arises in addition to the adiabatic one,
though the evolution of the isocurvature perturbations 
is model-dependent. 
Testing predictions of these inflation models will be the 
subject of the (future) precision cosmology by confronting 
with observation.
Statistics of the primordial fluctuations is very important
in such the test of the inflation models. In particular, the 
non-Gaussian nature of the primordial fluctuations is 
considered to be an important clue \cite{BKMR}, 
which stimulates the recent investigations on the higher order 
cosmological perturbations 
\cite{Enqvist,Vaihkonen,RSTA,RSTB,LMS,Longlois,LythA,LythB,LythC,Tomita,RSTC,Seery}.

The stochastic inflation was developed by Starobinsky and 
many works have been done \cite{Starobinsky,NNSa,NNSb,NNSc}. 
The stochastic approach to inflation is useful in 
investigating the statistics of the fluctuations \cite{SY,LMMR}. 
In the stochastic inflation,
the field dynamics coarse-grained over the horizon size
is described by the Langevin equation, and the field 
value of the inflaton is the probability variable.
Instead of solving the Langevin equation, 
we may solve the corresponding Fokker-Planck equation, 
which describes the evolution of the probability function $P(\phi)$. 
It is known that the scaling approximation is useful to 
describe the evolution of the probability function 
during the slow-roll period \cite{Suzuki}. 
With the use of the scaling approximation, the probability 
function can be obtained analytically \cite{MOL}. 
The stochastic approach to the multi-field inflation has been
considered in \cite{MMOL,GBW,Amendola}.
The probability function for the multi-field inflation was 
constructed using the scaling approximation in an analytic 
manner by Amendola \cite{Amendola}. However, the constructed 
probability function does not have the symmetry in  
exchanging the field parameters. 
Such the symmetry is desired for a consistency of the theory. 
As an extension of this previous work, we develop a useful 
analytic formula for the probability function of the 
multi-field stochastic inflation using the scaling approximation.

The organization of the present paper is as follows:
In section 2, we summarize the stochastic approach to the 
multi-field inflation. Then we obtain the probability
distribution function using the scaling approximation 
as an extension of the work by Amendola \cite{Amendola}. 
In section 3, based on the probability function, we 
investigate the non-Gaussian feature of the scalar fields 
during the inflation by decomposing the scalar fields 
into the adiabatic and the isocurvature components.
In section 4, we mention about connecting our result
with observation. Section 5 is devoted to summary and 
conclusions. Throughout this paper we use the unit $c=\hbar =1$, 
and follow the convention $(+,-,-,-)$. 

%----------------------------------------------------------
\section{Formulation}
%----------------------------------------------------------
\def\tlambda{{\tilde\lambda}}
We consider the chaotic inflation model with the two scalar fields 
$\phi $ and $\chi $ with the Lagrangian density
\begin{eqnarray}
  {\cal L}= {1\over 2}\nabla^\mu\phi\nabla_\mu\phi-V_1(\phi)
           +{1\over 2}\nabla^\mu\chi\nabla_\mu\chi-V_2(\chi),
\end{eqnarray}
where $V_1(\phi)$ and $V_2(\chi)$ are the potential, 
\begin{eqnarray}
  &&V_1(\phi)={\lambda_1\over 2n}\sigma^4
\left({\phi\over \sigma}\right)^{2n},
  \\
  &&V_2(\chi)={\lambda_2\over 2n}\sigma^4
\left({\chi\over \sigma}\right)^{2n}.
\end{eqnarray}
where $\lambda_1$ and $\lambda_2$ are the non-dimensional 
coupling constants, $n$ is the parameter, and we 
introduced $\sigma^{-2}=8\pi G$. 
During the slow-role regime, neglecting the 
kinetic terms,  the Friedman equation gives
\begin{eqnarray}
 H^2={V_1(\phi)+V_2(\chi)\over 3\sigma^2},
%={V(\phi,\chi)\over 3\sigma^2},
\end{eqnarray}
where $H=\dot a(t)/a(t)$ is the Hubble parameters, $a(t)$ 
is the scale factor, the dot denotes the derivative with 
respect to the cosmic time $t$.

We adopt the stochastic approach to the inflation. 
The basic equation, the Langevin equation, is obtained
by coarse-graining the field, averaging over the short 
wave length modes of the evolution equation of the field 
operator. For the two-field inflation, we have 
\begin{eqnarray}
  &&\dot\phi={-V_1,{}_\phi\over 3H}+g\eta_\phi
\label{LA}
\\
  &&\dot\chi={-V_2,{}_\chi\over 3H}+g\eta_\chi,
\label{LB}
\end{eqnarray}
where 
\begin{eqnarray}
g={H^{3/2}\over \sigma^{1/2}},
\end{eqnarray} 
and $\eta_\phi$ and 
$\eta_\chi$ describes the stochastic noises, which have
the nature of the Morkovian Gaussian noise with 
zero mean and the correlation
\begin{eqnarray}
&&\left< \eta_i (t)\right>=0, \\
&&\left< \eta_i (t)\eta_j (t')\right>=\delta_{ij}
  \frac{\sigma }{4\pi ^2}\delta (t-t'),
\end{eqnarray}
where $\eta_i$ and $\eta_j$ denote $\eta_\phi$ and $\eta_\chi$.
Thus the long-range scalar fields during the slow-roll regime
are described by the Langevin equations with the 
classical drift force $-V_1,{}_\phi/3H$ or $-V_2,{}_\chi/3H$
and the noise whose amplitude is in proportion to $g$. 
Note that, even if the field potentials consist of independent 
function of $\phi$ and $\chi$, the dynamics is coupled to each 
other because the Hubble parameter $H$ depends on the both 
fields. Note that we have assumed 
$\left< \eta_\phi (t)\eta_\chi (t')\right>=0$, 
based on the assumption that the fine-grained
component of the field $\phi$ and $\chi$ can
be treated as free in the leading approximation
\cite{MMOL,Amendola}.
Then, we may assume the Langevin equations for 
these field with the Gaussian noise with the 
correlation $\left< \eta_\phi (t)\eta_\phi (t')\right>
 =\left< \eta_\chi (t)\eta_\chi (t')\right>\propto 
\delta (t-t')$, in a similar way to the single field 
stochastic inflation. However, this assumption 
might be oversimplified for the models with general 
coupling between $\phi$ and $\chi$. 

Here let us consider the background solution for the Langevin 
equation without the noise term,
\begin{eqnarray}
  &&\dot\phi_{cl}={-V_1,{}_\phi\over 3H},
\label{cLA}
\\
  &&\dot\chi_{cl}={-V_2,{}_\chi\over 3H}.
\label{cLB}
\end{eqnarray}
From these zero-noise field equations, we have
\begin{eqnarray}
\frac{d\phi_{cl} }{d\chi_{cl} }=
\frac{\lambda _1\phi_{cl}^{2n-1}}{\lambda _2\chi_{cl}^{2n-1}}, 
\end{eqnarray}
which gives the attractor solution, assuming that the 
classical paths converge to it after a short transient,
\begin{eqnarray}
  \phi_{cl}=L^{1/2(n-1)}\chi_{cl},
\end{eqnarray}
where $L=\lambda_2/\lambda_1$. In the present paper, we refer 
to the attractor solution as the classical solution. We will 
solve the Fokker-Planck equation corresponding to the Langevin 
equation, along the classical attractor trajectory. 
For the latter convenience, 
we introduce the parameter $\theta$ by
\begin{eqnarray}
  &&\cos\theta={\dot \phi_{cl}\over 
\sqrt{\dot\phi_{cl}^2+\dot\chi_{cl}^2}}
=(1+L^{1/(1-n)})^{-1/2}
\\
  &&\sin\theta={\dot \chi_{cl}\over 
\sqrt{\dot\phi_{cl}^2+\dot\chi_{cl}^2}}
=(L^{1/(n-1)}+1)^{-1/2}.
\end{eqnarray}

Our formulation relies on rewriting the Langevin equation
in term of the variables $r$ and $\Theta$, introduced
in the below, instead of $\phi$ and $\chi$. First, 
we define the variable $r$ by 
\begin{eqnarray}
  r=\phi\cos\theta+\chi\sin\theta. 
\end{eqnarray}
From the Langevin equation for $\phi$ and $\chi$, we may write
\begin{eqnarray}
  &&\dot r=
  {-\tlambda\sigma^3\over 3H}\left({r\over \sigma}\right)^{2n-1}
  +g(\cos\theta\eta_\phi+\sin\theta\eta_\chi),
\label{LC}
\end{eqnarray}
where $H$ should be regarded as
\begin{eqnarray}
  &&H^2={1\over 3\sigma^2}\left(V_1(r\cos\theta)+V_2(r\sin\theta)\right)
  ={\tlambda\sigma^2\over 6n} \left({r\over \sigma}\right)^{2n}
\label{apH}
\end{eqnarray}
with defined
\begin{eqnarray}
  &&\tlambda=\lambda_1\cos^{2n}\theta+\lambda_2\sin^{2n}\theta
  =(\lambda_1^{1/(1-n)}+\lambda_2^{1/(1-n)})^{1-n}. 
\end{eqnarray}
The approximation of equation (\ref{apH}) is similar 
to that in \cite{Amendola}. In the strict sense, 
$H$ is the function of $\phi$ and $\chi$, but, in 
equation (\ref{apH}),  $H$ is approximated as 
$H(\phi,\chi)=H(\phi_{cl}(r), \chi_{cl}(r))$. 
The other variable $\Theta$ is defined as
\begin{eqnarray}
  \Theta={-1\over 2(n-1)}\left[
  {1\over \lambda_1} \left({\phi\over \sigma}\right)^{2(1-n)}
 -{1\over \lambda_2 }\left({\chi\over \sigma}\right)^{2(1-n)}
\right]. 
\end{eqnarray}
It is easy to show that $\Theta$ satisfies
\begin{eqnarray}
  &&{\dot\Theta}=\left(
  {\eta_\phi\over V_1,{}_\phi} 
  - {\eta_\chi\over V_2,{}_\chi} 
\right)g\sigma^2.
\label{LD}
\end{eqnarray}
We consider (\ref{LC}) and (\ref{LD}),
instead of the original Langevin equations 
(\ref{LA}) and (\ref{LB}). 

It will be important to mention about the physical meanings
of the variable $r$ and $\Theta$. It is clear that $r$ is
the variable parallel to the classical trajectory. On the
other hand $\Theta$ can be regarded to represent the component 
perpendicular to the classical trajectory. It is worthy
to note that the cross correlation of the noise terms
in the Langevin equations $r$ and $\Theta$ is zero, i.e.,
\begin{eqnarray}
  \left<
  (\cos\theta\eta_\phi+\sin\theta\eta_\chi)
  \left({\eta_\phi\over V_1,{}_\phi} 
  - {\eta_\chi\over V_2,{}_\chi} 
\right)\right>=0.
\end{eqnarray}

Now we consider solving the Fokker-Planck equation corresponding 
to the Langevin equations for $r$ (and $\Theta$). 
To find the solution analytically, we adopt the scaling approximation,
which was done in reference \cite{MOL} for single inflaton models. 
In the slow-roll regime of inflation, the drift force dominates 
the dynamical evolution of the scalar fields, and the quantum 
noise is small. Then the coarse-grained probability 
distribution develops a sharp peak around the classical slow-roll solution. 
This is the basic background that the scaling approximation 
may be used to describe the behavior of the probability 
distribution in the slow-roll regime.
The Langevin equation for $r$ is 
similar to that for the single field inflaton model, 
and we follow the reference \cite{MOL} to find
the analytic solution with the technique of 
the scaling approximation. In Appendix A, 
the derivation of the probability function is described.
The analytic probability function can be written as
function of $\phi$ and $\chi$ as 
\begin{eqnarray}
  P_{sc}(\phi ,\chi )=\frac{1}{2\pi \sqrt{\upsilon_{11}\upsilon _{22}}}
  \exp\left[-{(\xi(\phi,\chi)-\xi_*)^2\over 2\upsilon _{11}}
  -{\Theta(\phi,\chi)^2\over 2\upsilon _{22}}\right]|J(\phi ,\chi )|,
\label{PSC}
\end{eqnarray}
where $J(\phi,\chi)$ is the Jacobian of the transformation, 
\begin{eqnarray}
|J|=|\xi _{,\phi }\Theta _{,\chi }-\xi_{,\chi }\Theta _{,\phi }|, 
\end{eqnarray}
and the variances are obtained as
\begin{eqnarray} \upsilon _{11}=
\frac{1}{16\pi ^2}\sqrt{3\over 2n{\tilde \lambda }}\left(\frac{r_*}{\sigma }\right)^{2-n}\left[1-\left(\frac{r_{cl}}{r_*}\right)^4\right] 
\hspace{1cm} (n\ge2)
\label{vone}
\end{eqnarray}
and 
\begin{eqnarray}
  \upsilon _{22}=
\left\{\begin{array}{ll}{\displaystyle
\frac{{\tilde \lambda }}{192\pi ^2}\left(\frac{1}{\lambda _1^2\cos^6\theta }+\frac{1}{\lambda _2^2\sin^6\theta }\right)\left[-\ln\left(\frac{r_{cl}}{r_*}
\right)\right] }& (n=2) \\ 
{\displaystyle
\frac{{\tilde \lambda }}{96\pi ^2n^2(n-2)}\left(\frac{1}{\lambda _1^2\cos^{2(2n-1)}\theta} +\frac{1}{\lambda _2^2\sin^{2(2n-1)}\theta }\right)}&\\
{\displaystyle
\times\left[\left
(\frac{r_{cl}}{\sigma }\right)^{2(2-n)}-\left(\frac{r_*}{\sigma }
\right)^{2(2-n)}\right]} &(n>2)
\end{array}\right. ,
\label{vtwo}
\end{eqnarray}
and $\xi$ is related to $r(\phi,\chi)$ by equation (\ref{trr}) and $\xi_*=\xi(r_{cl})$.

Here we mention about the boundary condition of the
probability distribution function. The reflecting 
boundary condition, with which the normalization 
of the probability is preserved, is adopted in usual. 
In this case, other reflective term should be added 
to the probability function. However, 
as long as we are concerning with the behavior of 
the probability function around the peak of the classical 
slow-roll solution, we may work with the solution 
(\ref{PSC}) practically. 

Figure 1 shows the contours of the probability function
to make an example of its time evolution. 
At the earlier time (left panel), the probability function 
is almost spherically distributed. At the middle time
(center panel), the probability function becomes rather 
broad, then it becomes elongated along the direction of
the classical path at the latter time (right panel). 
The behavior depends on the model parameters as well as 
the initial condition.

%%%%%%%%%%%%%%%%%%%%%%%%%%%%%%%%%%%%%%%%%%%%%%%%%%%%%%%%%%%%
\section{Non-Gaussian behavior of the probability function}
%%%%%%%%%%%%%%%%%%%%%%%%%%%%%%%%%%%%%%%%%%%%%%%%%%%%%%%%%%%%
\def\calN{{\cal N}}
In this section we investigate the evolution of the
probability function (\ref{PSC}), focusing on the 
non-Gaussian statistical nature. We start from 
introducing the variable $\Delta r$ and $\Delta s$ 
by
%\begin{eqnarray}\left(
%\begin{array}{cc}
%{\phi} \\
%{\chi} \\
%\end{array}
%\right)=\left(
%\begin{array}{cc}
%\phi _{cl} \\
%\chi _{cl} \\
%\end{array}\right)+
%\left(
%\begin{array}{cc}
%\cos \theta & -\sin \theta  \\
%\sin \theta & \cos \theta  \\
%\end{array}
%\right)\left(
%\begin{array}{cc}
%\{\Delta r} \\
%{\Delta s} \\
%\end{array}
%\right).
%\end{eqnarray}
%This can be written as 
\begin{eqnarray}\left(
\begin{array}{cc}
{\Delta r} \\
{\Delta s} \\
\end{array}
\right)=
\left(
\begin{array}{cc}
\cos \theta & \sin \theta  \\
-\sin \theta & \cos \theta  \\
\end{array}
\right)\left(
\begin{array}{cc}
{\phi-\phi_{cl}} \\
{\chi-\chi_{cl}} \\
\end{array}
\right).
\end{eqnarray}
We may regard $\Delta r$ ($\Delta s$) as the component that 
contributes to the adiabatic (isocurvature) perturbation 
\cite{Gordon,AGWS}. 
We rewrite the probability distribution function in terms 
of $\Delta r$ and $\Delta s$. The expansion can be done
in a straightforward manner. But the expression is rather long, 
then the result is given in Appendix B. 

Here let us introduce the $e$-folding number from a time $t$ 
to the final time of the inflation $t_f$,
\begin{eqnarray}
  N(t)=\int_{t}^{t_f} H(t')dt',
\label{defefold}
\end{eqnarray}
where $t_f$ is defined by 
$V_1(\phi_{cl})+V_2(\chi_{cl})
  =\left(\dot \phi_{cl}^2+\dot \chi_{cl}^2\right)/2$,
which gives $r_{cl}(t_f)^2=2n^2\sigma^2/3$. 
Instead of using $t$ (or $r_{cl}(t)$), we use $N$ as 
the time variable. The classical solution 
%$r_{cl}(r_{cl}/r_*)$
is related to $N$ by
\begin{eqnarray}
\left({r_{cl}(t)\over r_*}\right)^2
=\frac{n+6{N}}{n+6N_*}\equiv {\cal N}\le 1
\end{eqnarray}
for $n\ge2$, where $N_*=N(t_*)$. 

With the use of ${\cal N}$ (or $N$), the part of the 
probability function of $\xi$, equation (\ref{expxi}),  
is written
\begin{eqnarray}
  \exp\left[-{(\xi-\xi_*)^2\over 2\upsilon _{11}}\right]
  &=&\exp\Biggl[-R_n
  \Biggl\{\left({\Delta r\over r_{cl}}\right)^2
\nonumber\\
&& -\left(n-1+{n+2\over2}\calN^{2-n}\right)
  \left({\Delta r\over r_{cl}}\right)^3 
\nonumber \\
&& +\Biggl( 
  \frac{(n-1)(7n-3)}{12}+\frac{3(n-1)(n+2)}{4}
  \calN^{2-n}
\nonumber\\
&&+\frac{(n+2)(30-n)}{48}\calN^{2(2-n)} \Biggr)  
  \left(\frac{\Delta r}{r_{cl}}\right)^4
  \cdots\Biggr\}\Biggr],
\label{expxib}
\end{eqnarray}
and $R_n$ is
\begin{eqnarray}
R_n={96\pi^2 \over \tlambda}
\left(\frac{3}{2n}\right)^{n}n^2({\cal N}^{-2} - 1 )^{-1}(n+6{N})^{-n}
\label{Rnb}
\end{eqnarray}
from (\ref{Rn}). On the other hand, the part of the probability function of $\Theta$ 
is also written in the similar way, i.e., equation (\ref{exptheta}), 
with
\begin{eqnarray}
  S_n={96\pi^2 \over \tlambda}\times\left\{
\begin{array}{ll}
{\displaystyle
\frac{9}{32}(1+3{N})^{-2}(-\ln {\cal N} )^{-1}} & (n=2) 
\nonumber\\
{\displaystyle
\frac{(n-2)n^2}{2}\left(\frac{3}{2n}\right)^n(n+6{N})^{-n}
  (1- {\cal N}^{n-2} )^{-1}} & (n>2)
\end{array} \right.\\
\label{Snb}
\end{eqnarray}

From equations (\ref{Rnb}) and (\ref{Snb}) we easily have
\begin{eqnarray}
\frac{S_n}{R_n}=\left\{
\begin{array}{ll}
{\displaystyle
\frac{1}{2}({\cal N}^{-2}-1)(-\ln {\cal N} )^{-1}} & (n=2) \\
{\displaystyle
\frac{n-2}{2}({\cal N}^{-2}-1)(1-{\cal N}^{n-2})^{-1} }& (n>2) 
\end{array} \right. 
\label{SR}
\end{eqnarray}
and $S_n/R_n\ge 1$ (see Figure 2). Because the variance of 
$\Delta r$ and $\Delta s$ is in proportion to $1/R_n$
and $1/S_n$, respectively, then it can be read that  
the amplitude of the fluctuation of the isocurvature 
component is smaller than that of the adiabatic component. 
This is consistent with the behavior of the probability 
function shown in Figure 1. 

We next consider the deviation of the probability function from 
the Gaussian statistics, which is expressed by the terms in 
proportion of $\Delta r^3$, $\Delta s^3$, and the other higher 
order terms. 
To investigate the deviation of the probability distribution 
function from the Gaussian statistics quantitatively, we 
first define the ratio
\begin{eqnarray}
{\cal T}\equiv 
{P_{sc}(+\Delta r,+\Delta s)\over P_{sc}(-\Delta r, -\Delta s)}.
%\frac{P_{sc}(\phi _{cl}+\Delta \phi (\Delta r,\Delta s),
%\chi _{cl}+\Delta \chi(\Delta r,\Delta s) )}
%{P_{sc}(\phi _{cl}-\Delta \phi (\Delta r,\Delta s),
%\chi _{cl}-\Delta \chi(\Delta r,\Delta s) )}.
\end{eqnarray}
Thus ${\cal T}$ represents the asymmetry of the probability function,
which can be related to the dimensionless skewness (see below). 
In the limit of the Gaussian statistics, ${\cal T}-1$
approaches to zero. This is a very simple but useful estimator of the 
local skewness around the Gaussian peak introduced in reference \cite{Mine}. 
{}From our probability function, we have the expression of the leading
order 
\begin{eqnarray}
{\cal T}-1&\approx &2\Biggl[
\left(n-1+\frac{n+2}{2}{\cal N}^{(2-n)/2}\right)R_n
  \left({\Delta r\over r_{cl}}\right)^3 - (2n-1){\Gamma_n}S_n
  \left({\Delta s\over r_{cl}}\right)^3
\nonumber\\
  &&\hspace{4.5cm}+ 2(2n-1)S_n\left({\Delta r\over r_{cl}}\right)
  \left({\Delta s\over r_{cl}}\right)^2
\Biggr],
\label{NONG}
\end{eqnarray}
for $n\ge 2$, where $\Gamma_n$ is defined by (\ref{gamman}).
Our result of the adiabatic component with $n=2$ reproduces the
result in reference \cite{Mine}.

We can read a few characteristic features 
of the skewness from expression (\ref{NONG}). 
First, the evolution of the skewness, in general, depends 
on the parameters of the potential. It also depends 
on the initial time $N(t_*)$, which we have adopted to 
solve the Fokker-Planck equation.  
This is unavoidable in our scheme.
Let us consider the skewness of the adiabatic component, 
the first term of the right hand side (r.h.s.) of (\ref{NONG}). 
To estimate the skewness quantitatively, we consider the
first term at the 
%$1-\sigma$ 
one sigma deviation with respect to 
$\Delta r$, with replacing $\Delta r/r_{cl}$ by $1/\sqrt{2R_n}$,
\begin{eqnarray}
 A_R= \left(n-1+\frac{n+2}{2}{\cal N}^{(2-n)/2}\right)
  {1\over \sqrt{2R_n}}.  
\end{eqnarray}
In the similar way, considering the second term of (r.h.s.) 
of (\ref{NONG}) 
%at the $1-\sigma$ 
at the one sigma deviation with respect to 
$\Delta s$, with replacing $\Delta s/r_{cl}$ by $1/\sqrt{2S_n}$,
we define
\begin{eqnarray}
 A_S=- (2n-1){\Gamma_n}{1\over \sqrt{2S_n}}.  
\end{eqnarray}
Figure 3 plots $A_R$ and $A_S$ as function of $N$ 
for the models $\lambda_1=\lambda_2=10^{-12}$,
and $\lambda_1=10^{-12}$ and $\lambda_2=10^{-13}$. 
For the model $\lambda_1=\lambda_2$, we have $A_S=0$
because $\Gamma_n=0$. Here we extensively used our result to the case 
$n=1.5$. We see that $A_S$ is larger than $A_R$ for the model
$\lambda_1=10^{-12}$ and $\lambda_2=10^{-13}$. 
Figure 4 plots the same as Figure 3 but with $n=2$. 
Similarly, Figures 5 and 6 are  same as Figure 3 
but with $n=3$ and $n=4$, respectively.
{}From Figures 3-6, we see that the non-Gaussian 
estimator $A_R$ and $A_S$ becomes larger as $n$ becomes larger, 
in general, depending on $\tilde \lambda$. 
But the ratio of $A_S$ to $A_R$ becomes large as 
$n$ becomes small. This comes from the fact that the 
skewness of the isocurvature component $A_S$ is in 
proportion to $\Gamma_n$. 
Figure 7 plots the ratio $A_S/A_R$ at a fixed time as 
function of $L$ for various models of $n$. 
This result demonstrates that the skewness of the isocurvature 
component is independent of the adiabatic component, and can be
large compared to the adiabatic one. The asymmetry in the two
potentials enhances the non-Gaussianity of the isocurvature 
component. In reference \cite{Bernardeau}, the importance of 
the non-Gaussianity of the isocurvature component was discussed in 
a different context of the investigation of multi-field inflation.

The third term of the r.h.s. of (\ref{NONG}) represents the 
mixing of the isocurvature and the adiabatic components. 
This suggests that the adiabatic and isocurvature 
components may have a correlation at nonlinear order 
in general, even if it have no correlation at the 
linear order \cite{BMR}.

Similar to the skewness , the terms in proportion 
to $\Delta r^4$ and $\Delta s^4$ represent the local kurtosis 
of the probability function. 
It can be read that the coefficient of $\Delta s^4$ 
depends on $\Gamma_n$ but the coefficient of $\Delta r^4$ 
does not. This means that, similar to the case of the skewness,  
the kurtosis of the isocurvature component is independent of
the adiabatic component and can be large compared to it.  

%%%%%%%%%%%%%%%%%%%%%%%%%%%%%%%%%%%%%%%%%%%%%%%%%%%%%%%%%%%
\section{Connecting the result with observation}
%%%%%%%%%%%%%%%%%%%%%%%%%%%%%%%%%%%%%%%%%%%%%%%%%%%%%%%%%%%
\def\fnl{{f_{NL}}}
Here let us briefly mention about the detectability of the skewness
and kurtosis. 
As mentioned in section 1, many works
related to the non-Gaussian nature of the primordial 
fluctuations have been done. 
The useful quantity $\fnl$ 
is introduced to characterize the non-Gaussianity of the
gravitational potential $\Phi$, by expanding it as \cite{KS} 
\begin{eqnarray}
  \Phi=\varphi+\fnl(\varphi^2-\left<\varphi^2\right>)+{\cal O}(\fnl^2),
\end{eqnarray} 
where $\varphi$ is a zero-mean Gaussian probability variable.
Here assume $\varphi$ and $\Phi$ are the variables with one 
degree of freedom, it is easy to construct the probability
distribution function of $\Phi$,
\begin{eqnarray}
  P(\Phi)\propto{}\exp\left[
  -{{\Phi^2-2\fnl\Phi^3}\over 2\left<\Phi^2\right>}
\right],
\label{PPhi}
\end{eqnarray} 
neglecting the terms proportional to $\fnl^2$. 
\footnote{In deriving this expression we redefined
$\Phi+\fnl\left<\Phi^2\right>$ by $\Phi$ to eliminate the 
term proportional to $\Phi$.} 
Then we define the local skewness for the probability 
function $P(\Phi)$, in the similar way 
to the previous section, by
\begin{eqnarray}
  {\cal T}_\Phi-1={P(\sqrt{\left<\Phi^2\right>}) \over P(-\sqrt{\left<\Phi^2\right>})} -1
  \simeq
  {2\fnl \left<\Phi^2\right>^{1/2}}.
\end{eqnarray} 
We can write ${\cal T}_\Phi-1$ in terms of 
the dimensionless skewness defined by 
$\left<\Phi^3\right>/\left<\Phi^2\right>^{3/2}$.
Approximating the probability function as
\begin{eqnarray}
\exp\left[  -{{\Phi^2-2\fnl\Phi^3}\over 2\left<\Phi^2\right>}\right]
\simeq \left(1+f_{NL}{\Phi^3\over\left<\Phi^2\right>}\right)
\exp\left[ -{\Phi^2\over 2\left<\Phi^2\right>}\right],
\end{eqnarray} 
we have
\begin{eqnarray}
  {\left<\Phi^3\right>\over\left<\Phi^2\right>^{3/2}}
  \simeq 15 \fnl \sqrt{\left<\Phi^2\right>}={15\over 2} ({\cal T}_\Phi-1).
\end{eqnarray} 

In general, the evolution of fluctuations depends on each model.
However, we here introduce the assumption that the dimensionless 
skewness preserves, that is $|{\cal T}_\Phi-1|\simeq|{\cal T}-1|$,  
even if the amplitude of the fluctuations may change during the evolution 
of the universe. Based on this assumption our result can be
connected with observation. In this case the predicted local 
skewness is related to $f_{NL}$ as  
\begin{eqnarray}
  |{\cal T}-1|={2\fnl \sqrt{\left<\Phi^2\right>}}
{\hspace{1cm}}
{\rm or}
{\hspace{1cm}}
  {|{\cal T}-1|\over \sqrt{\left<\Phi^2\right>}} ={2\fnl }.
\end{eqnarray} 
If we adopt $ \sqrt{\left<\Phi^2\right>}=10^{-5}$, the vertical axis of 
Figures 3-6 can be read as ${2\fnl}$. 
%Here we assume that the probability function at the $e$-folding $N$ 
%describes the probability function of the fields coarse-grained 
%over the scale that the horizon crossing occurs at that time. 

The detectable minimum value of $f_{NL}$ is discussed in references 
\cite{KS}, $f_{NL}=20$ and $5$ for WMAP and Planck data, respectively. 
Thus the non-Gaussianity in our model with $n\ge3$ might be detectable.
However, this is based on the rough estimation because we have not 
strictly considered the observational constraint on the model parameters 
from the spectrum index and the amplitude of the perturbation. 
For definite conclusions we need incorporate these constraints. 

Similar to the skewness, we can estimate the local kurtosis,
%$\left<\Phi^4\right>/\left<\Phi^2\right>^2$, 
being of order
${\cal O}(\left<\Phi^2\right>)$. This is smaller than
the skewness by a factor of $10^{-5}$. 

%%%%%%%%%%%%%%%%%%%%%%%%%%%%%%%%%%%%%%%%%%%%%%%%%%%%%%%%%%%
\section{Summary and conclusions}
%%%%%%%%%%%%%%%%%%%%%%%%%%%%%%%%%%%%%%%%%%%%%%%%%%%%%%%%%%%
In the present paper, we have investigated the statistical nature
of the multi-field inflation with the stochastic approach, 
in which the evolution of the scalar fields is described by the 
Langevin equations. To solve the corresponding Fokker-Planck equation
analytically, we adopt the technique of the scaling approximation.
This was first considered in \cite{Amendola}. We have improved 
the previous investigation requiring the probability function 
to have the symmetry in exchanging the field parameters. 
We have investigated the non-Gaussian nature of the probability 
function by decomposing the fields into the adiabatic and the 
isocurvature components. 

We have introduced the local skewness as an estimator of the 
non-Gaussianity, and have found the followings: (1) The amplitude 
of the skewness is determined by the combination of the 
coupling constant $\lambda_1$, $\lambda_2$ and $n$.  
The amplitude of the skewness of the adiabatic component is 
determined by the 
combination of the coupling constant $\tilde \lambda$ and $n$, 
while that of the isocurvature component depends on 
$\tilde \lambda$, $n$ and the ratio of the coupling constant 
$\Gamma_n$. Then the skewness of the 
isocurvature component can be large compared with that of the 
adiabatic one due to the factor $\Gamma_n$. 
(2) The similar feature can be seen in the kurtosis of
the adiabatic and isocurvature components. (3) The adiabatic 
and isocurvature components may have correlation at the nonlinear 
order in the skewness and kurtosis, even if there is no 
correlation at the linear order. 

We have considered the connection between our result and
observation, based on the assumption that the dimensionless 
skewness of fluctuation is preserved. This argument shows that the skewness
will be useful to constrain the potential parameter.
However the assumption is not trivial and 
there remains room for more discussions for precise 
comparison with observation. 
Furthermore, in the present paper, we have only 
worked with the simple chaotic inflation models with the power-low 
potentials. Investigation of multi-field inflation with 
more general potentials will be an interesting subject,
including the confrontation with cosmic microwave anisotropy
observation.
Testing the robustness of the approximation adopted here
will also be required as a future investigation.

\ack
This work is supported in part by Grant-in-Aid for Scientific 
research of Japanese Ministry of Education, Culture, Sports, 
Science and Technology, No.15740155. We thank B. A Bassett and 
anonymous referee for useful comments, which helped improve the 
manuscript. 
We also thank M. Sasaki for useful conversation 
related to the topic in the present paper.

\vspace{5cm}
%\newpage
%\newpage
\begin{figure}[b]
\begin{center}
\epsfbox{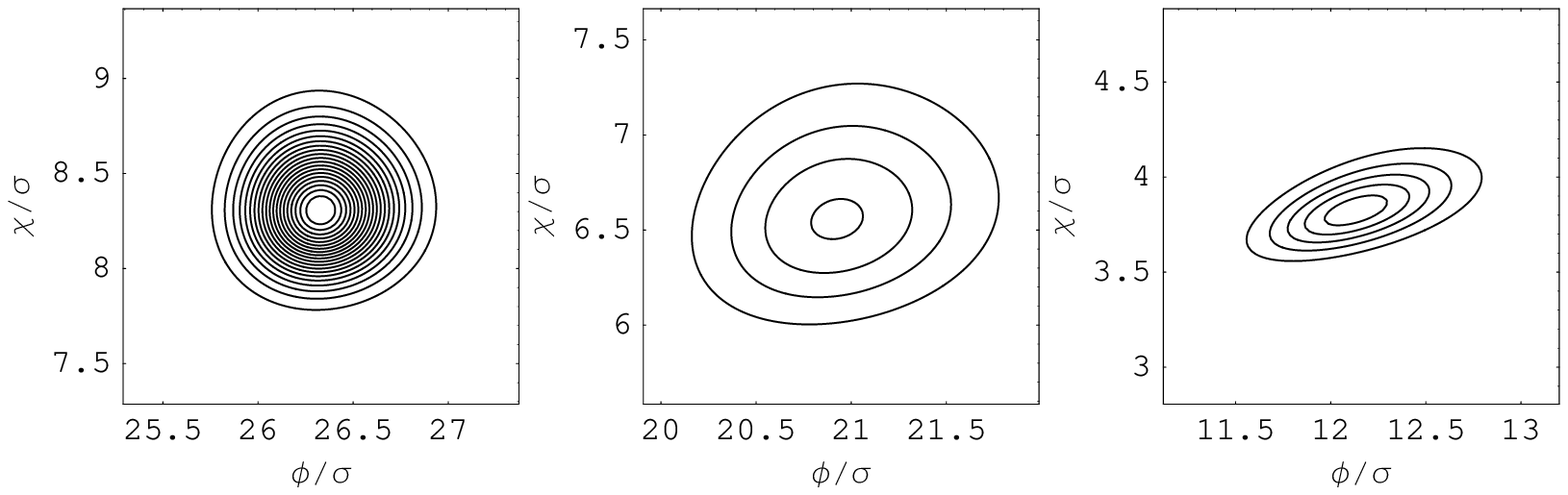}
\end{center}
\caption{Contours of the probability distribution function. 
Here the potential with $\lambda _1=10^{-5}$, 
$\lambda _2=10^{-4}$, and $n=2$ is considered. 
The initial time is chosen $N_*=100$ and 
the left, middle and right panels show
the contour at the time of the e-folding $N=95$, 
$60$, and $20$, respectively. (See equation (28) for the 
definition of $N$)
\label{labelA}}
\vspace{2.5cm}
%\end{figure}
%\begin{figure}
\begin{center}
\epsfbox{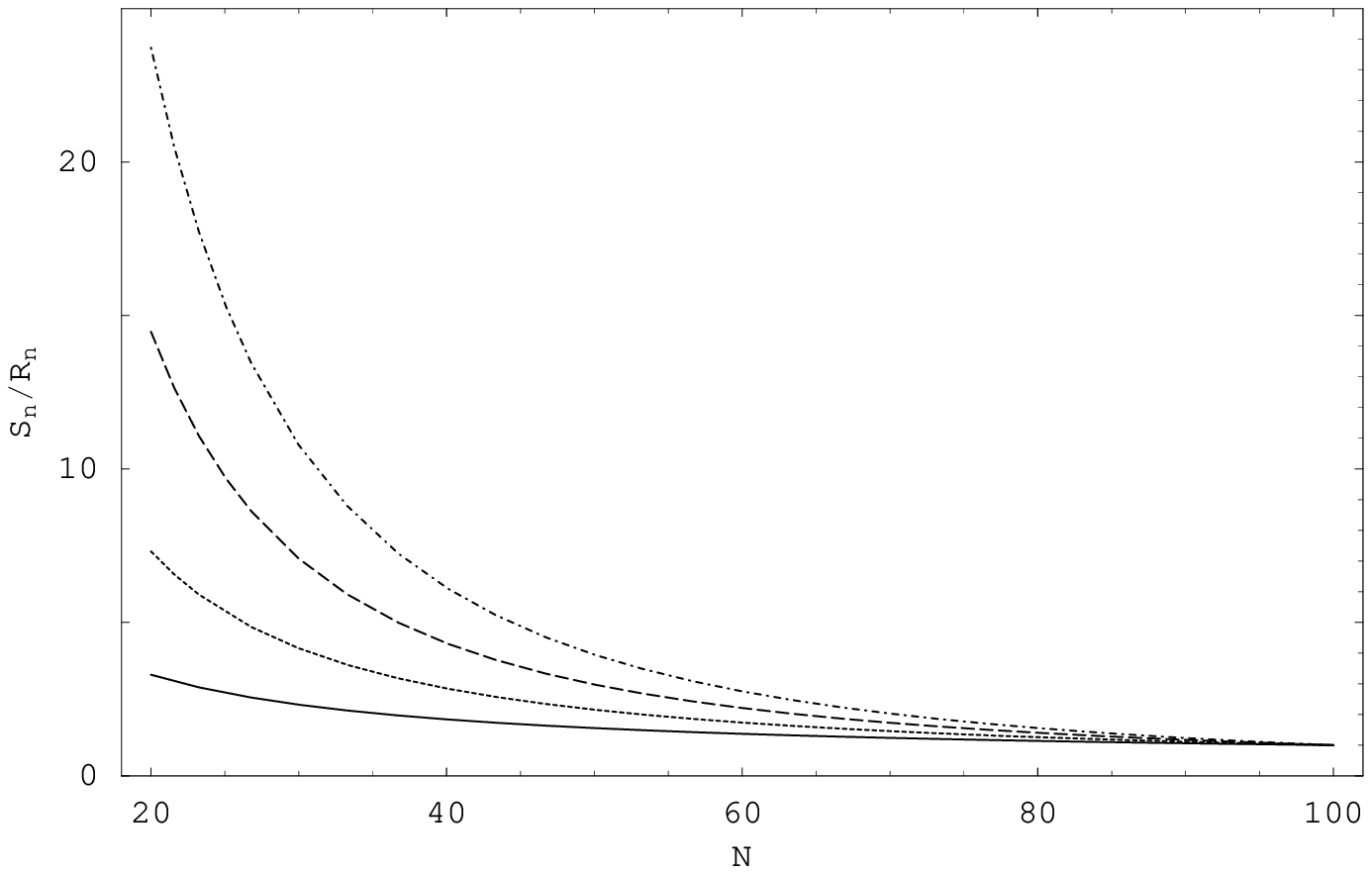}
%\epsfbox{amplitude.eps}
\end{center}
\caption{The ratio $S_n/R_n$ is plotted as function of $N$. 
The initial time is chosen $N_*=100$. The parameter of 
the potential $n$ is chosen $n=1.1,~2,~3,~4$ from bottom 
to top. Here we extensively used our result to the case $n<2$.
\label{labelB}}
\vspace{1cm}
\end{figure}

\newpage
\begin{figure}
\begin{center}
\epsfbox{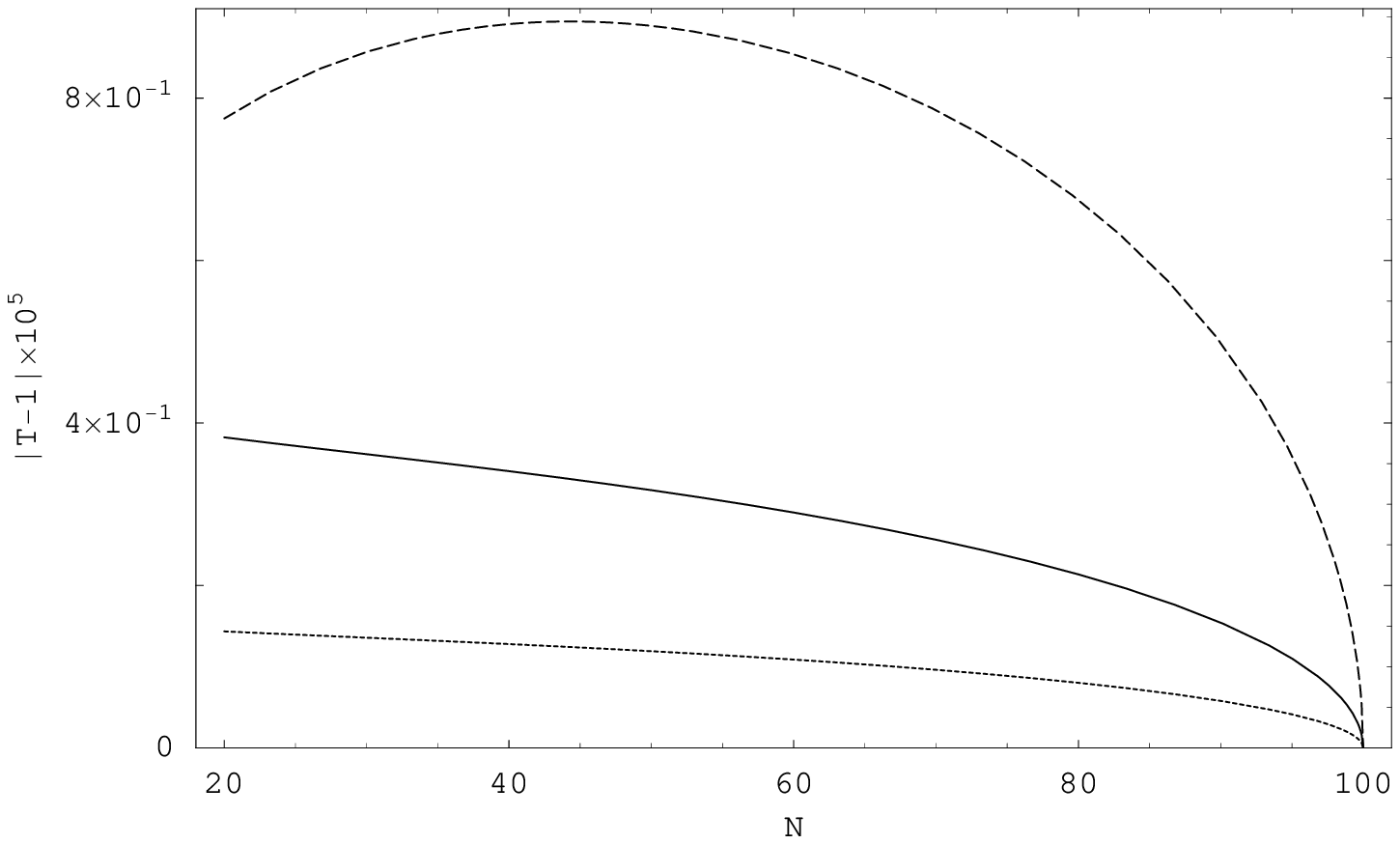}
%\epsfbox{non-g15.eps}
\end{center}
\caption{$A_R$ and $A_S$ as function of $N$. 
The dotted (dashed) curve plots $A_R$ ($A_S$) for the model 
with the parameters $\lambda_1=10^{-12}$ and $\lambda_2=10^{-13}$, 
while the solid curve plots $A_R$ with $\lambda_1=\lambda_2=10^{-12}$.
Here we adopted $N_*=100$ and $n=1.5$. 
\label{labelC}}
%\vspace{1cm}
%\end{figure}
%
%\newpage
%\begin{figure}
\begin{center}
\epsfbox{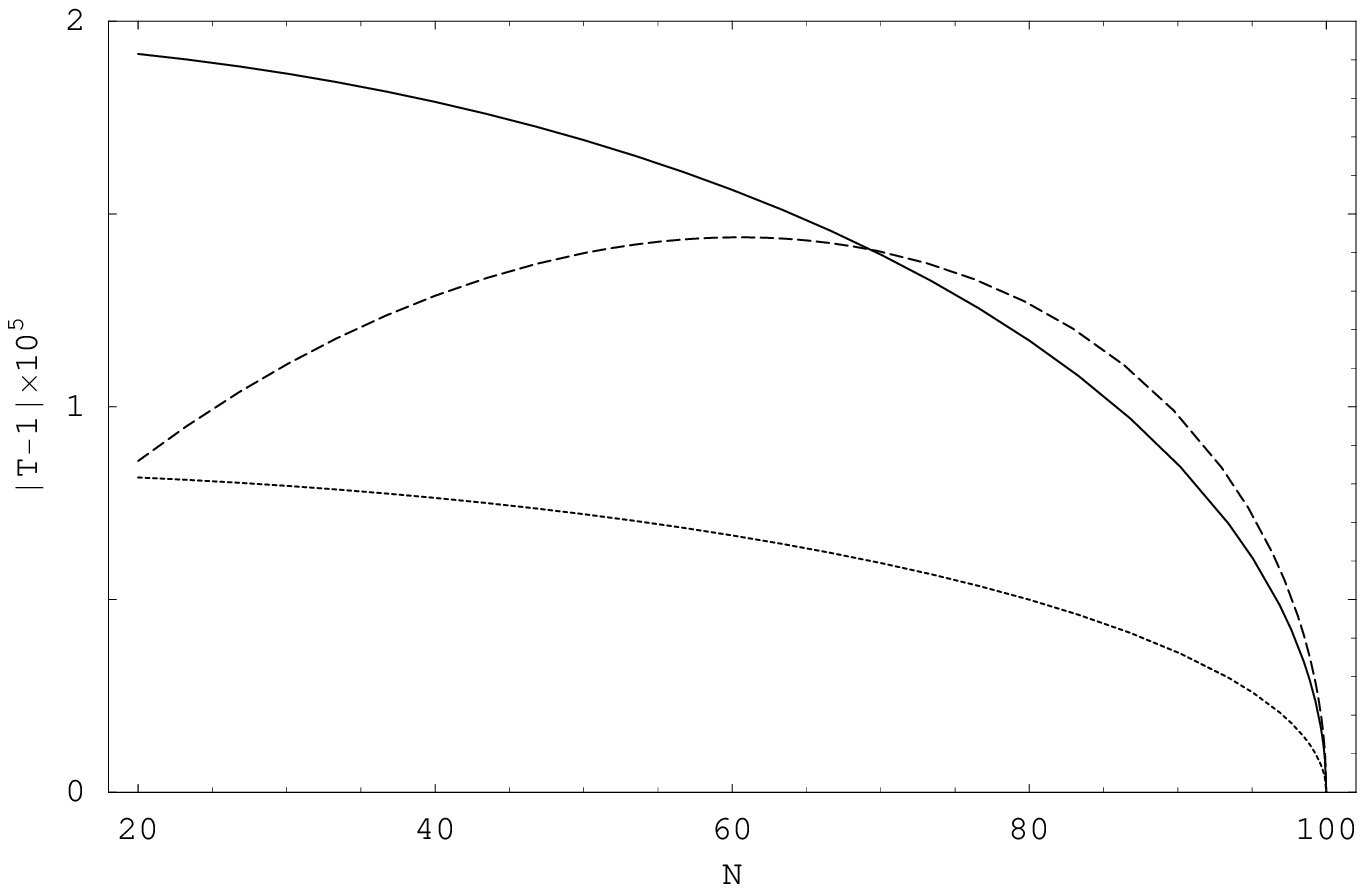}
%\epsfbox{non-g2.eps}
\end{center}
\caption{Same as figure 3, but with $n=2$.
\label{labelD}}
\vspace{0.5cm}
\end{figure}
%

%\newpage
\begin{figure}
\begin{center}
\epsfbox{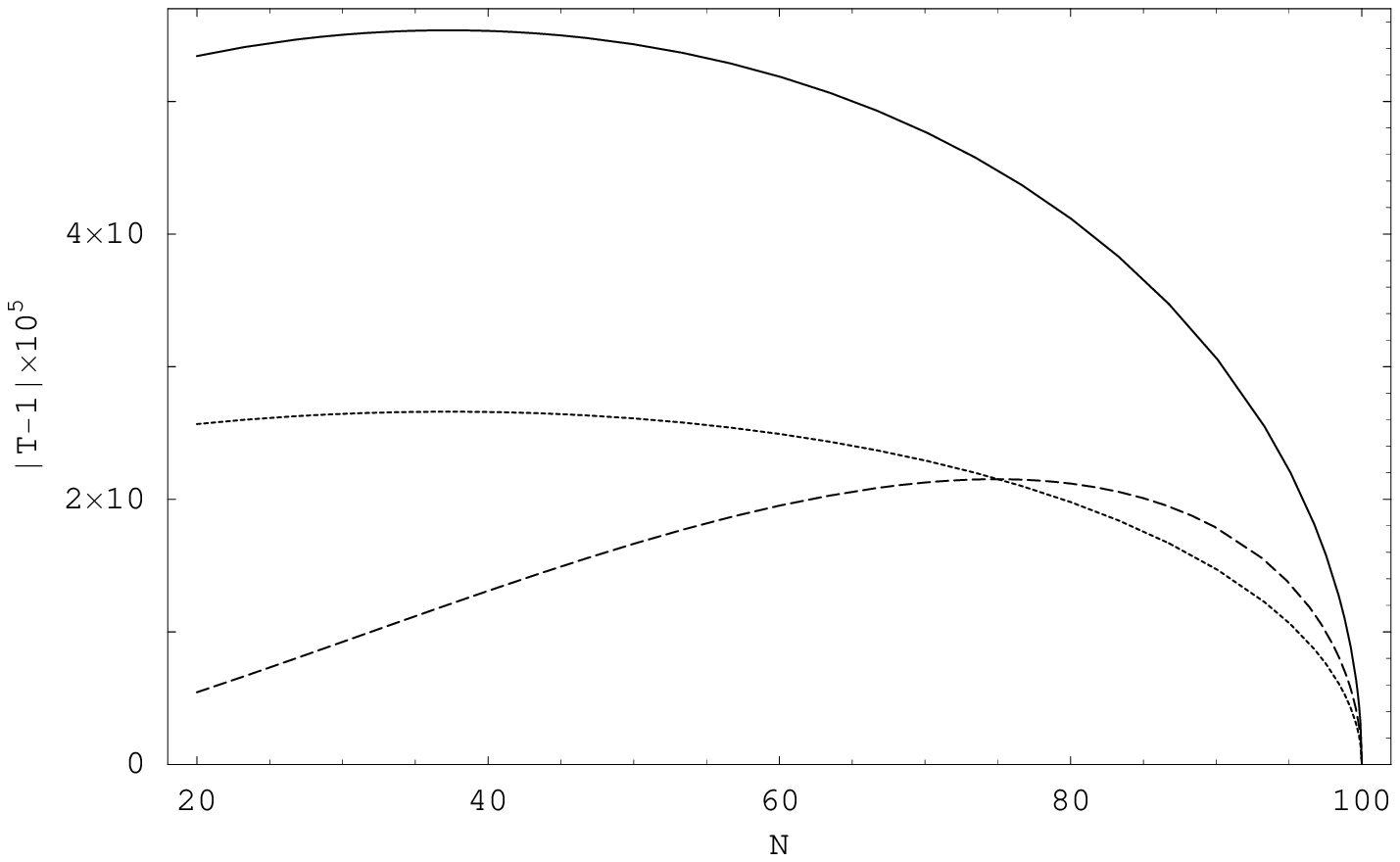}
%\epsfbox{non-g3.eps}
\end{center}
\caption{Same as figure 3, but with $n=3$.
\label{labelDD}}
\vspace{0.5cm}
%\end{figure}
%
%
%\newpage
%\begin{figure}
\begin{center}
\epsfbox{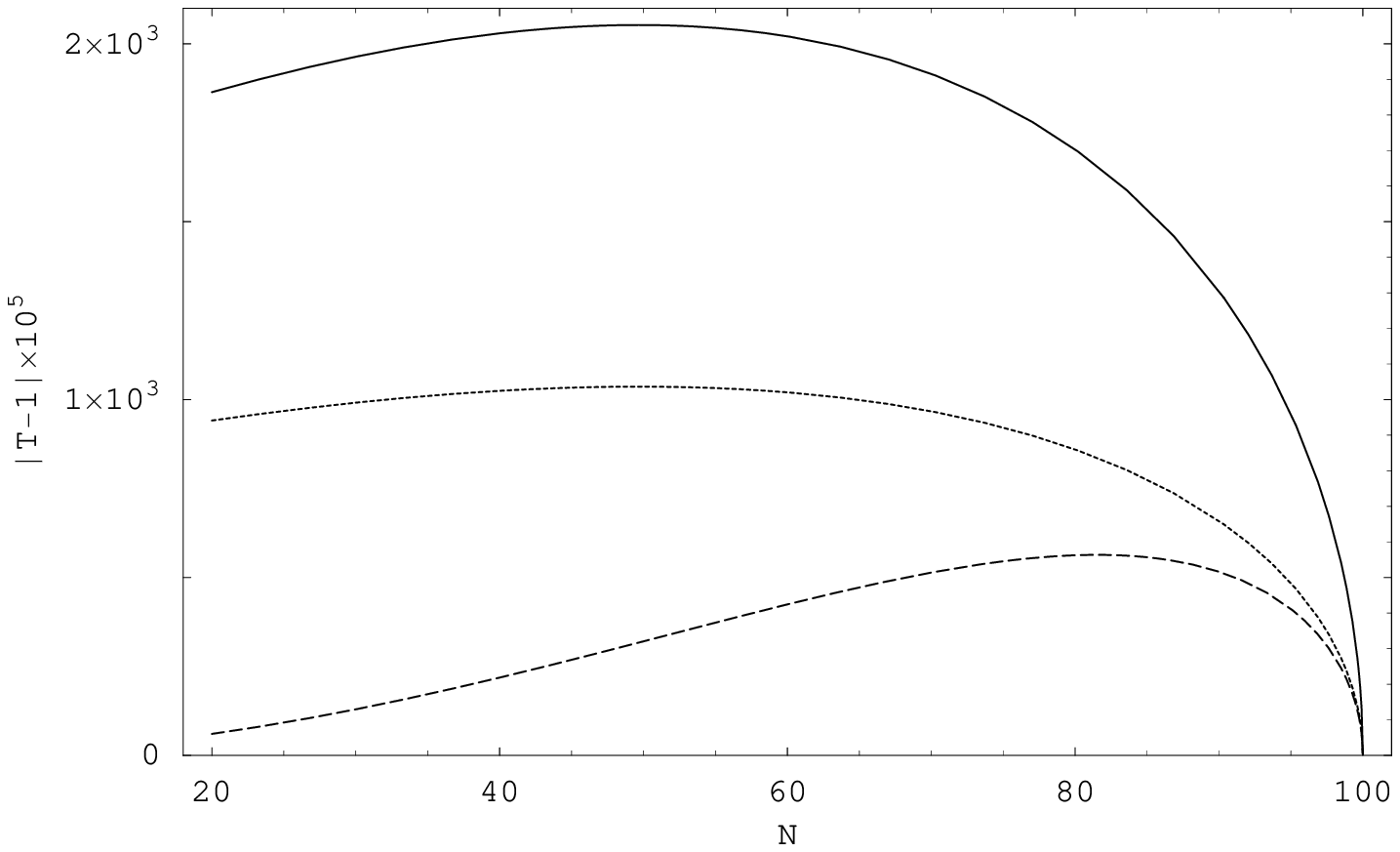}
%\epsfbox{non-g4.eps}
\end{center}
\caption{Same as figure 3, but with $n=4$
\label{labelE}}
\vspace{1.0cm}
\end{figure}
%

%\newpage
\begin{figure}
\begin{center}
\epsfbox{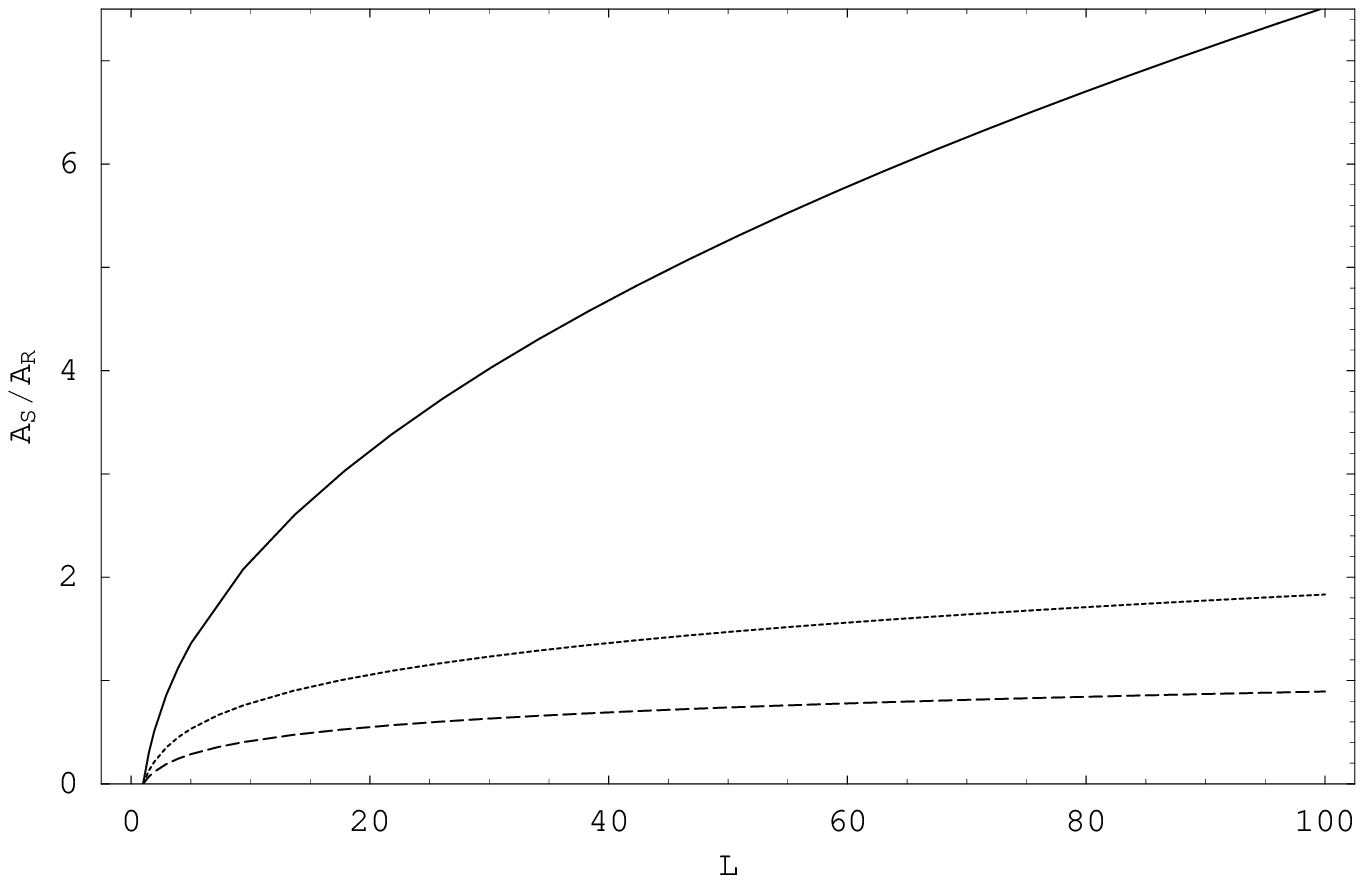}
%\epsfbox{non-g_l.eps}
\end{center}
\caption{The ratio $A_S/A_R$ at the time $N=60$ as function of $L$ for 
the models $n=2,~3,~4$ (from top to bottom). Here we fixed 
$\lambda_1\lambda_2=10^{-24}$, and $N_*=100$ is adopted.  
\label{labelF}}
\vspace{1.5cm}
\end{figure}
% Figure

%\newpage
\appendix
%%%%%%%%%%%%%%%%%%%%%%%%%%%%%%%%%%%%%%%%%%%%%%%%%%%%%%%%%%%%%%%%%%%%
\section{Scaling transformation and probability function}
%%%%%%%%%%%%%%%%%%%%%%%%%%%%%%%%%%%%%%%%%%%%%%%%%%%%%%%%%%%%%%%%%%%%
To find the scaling solution of equation (\ref{LC}), 
two transformation on the variable is performed to 
yield a Langevin equation with drift term vanished. 
The first transformation is defined by
\begin{eqnarray}
  r\to z=\int_r^{r_\infty}\frac{dr'}{g(r')},
\end{eqnarray}
where $r_\infty$ is an arbitrary constant, but we here chose 
$r_\infty=\infty$ to get $z$ with the same sign of $r(>0)$. 
%; the domain of $z$ is limited to either $r>0$ or $r<0$).
In terms of this new variable, the Langevin equation is
\begin{eqnarray}
% {\dot z}=-\frac{\sigma ^{1/2}}{3H^{5/2}}\frac{\partial V(r)}
% {\partial r}+\cos \theta \eta _{\phi }+\sin \theta\eta _{\chi}.
  &&\dot z=
  {\tlambda\sigma^{7/2}\over 3H^{5/2}(r(z))}
  \left({r(z)\over \sigma}\right)^{2n-1}
  -(\cos\theta\eta_\phi+\sin\theta\eta_\chi).
\label{LE}
\end{eqnarray}
Thus, with the first transformation, the multiplicative 
($r$-dependent noise) process reduces to the additive 
(constant noise) process, and
the variable $z$ is referred to as the constant-diffusion 
variable. 
Then we perform the second mapping,
requiring that the resultant Langevin equation 
has no drift term,
\begin{eqnarray}
  z\to \xi =F^{-1}[e^{\Lambda t}F(z)].
\end{eqnarray}
The function $F$ is found in the form, 
\begin{eqnarray}
  F(z)=\exp \left[-\int_{z_*}^z\frac{\Lambda }{D_z}dz'\right],
\end{eqnarray}
where $D_z$ is the drift term of equation (\ref{LE}),
\begin{eqnarray}
  D_z={\tlambda\sigma^{7/2}\over 3H^{5/2}(r(z))}
  \left({r(z)\over \sigma}\right)^{2n-1},
\end{eqnarray} 
and $\Lambda =-D_z(z_*)$ is the separation constant and $z_*$ 
is chosen so that $\partial F(z_*)/\partial z=1$. 
Through these transformation, we obtain
\begin{eqnarray}
\xi(r)=\left\{
\begin{array}{ll}
{\displaystyle {1\over 2}\left({12\over\tlambda}\right)^{3/4}
  \left({r\over \sigma}\right)^{-2}
  \left({r_{cl}\over r_*}\right)^{2}} &(n=2) \\
{\displaystyle {2\over 3n-2}\left({6n\over\tlambda}\right)^{3/4}}&\\
{\displaystyle \times\left[ \left({r\over \sigma}\right)^{2-n}
  -\left({r_{cl}\over \sigma}\right)^{2-n}
  +\left({r_*\over \sigma}\right)^{2-n}
\right]^{(3n-2)/(2n-4)}} &(n>2) 
\end{array}\right. 
\label{trr}
\end{eqnarray}
where $r_*$ is the initial value of the classical solution, and 
we used the classical solution $r_{cl}(t)$, defined as
\begin{eqnarray}
{r_{cl}\over r_*}=
\left\{
\begin{array}{ll}
{\displaystyle \exp\Biggl[-2\sqrt{\tlambda\over 3}\sigma t\Biggr] }& (n=2) \\
{\displaystyle \Biggl[1+(n-2)\left({r_*\over \sigma}\right)^{n-2}\sqrt{2n{\tilde \lambda }\over 3}\sigma t\Biggr]^{1/(2-n)} }& (n > 2) 
\end{array}\right. ,
\end{eqnarray}
which is given by solving equation (\ref{LC}) neglecting the noise term.
Now the Langevin equation for $\xi$ is 
 \begin{eqnarray}
{\dot \xi }=-\left(\frac{\xi }{z}\right)^{(n+2)/(3n-2)}
  (\cos \theta \eta_{\phi} +\sin \theta\eta_{\chi }).
\end{eqnarray}
We follow the standard choice of the initial condition for the 
probability distribution, the delta-function distribution at 
the initial time. Furthermore, assuming the limit of small diffusion, 
we adopt the Gaussian form of the probability distribution function
in the scaling approximation
\begin{eqnarray}
 P(\xi)\propto \exp\left[-{(\xi-\xi_*)^2\over 2\upsilon_{11}}\right],
\end{eqnarray}
where $\upsilon_{11}$ is given by expression (\ref{vone}), and $\xi_*=\xi(r_{cl})$. 

On the other hand the Langevin equation for the variable $\Theta$ 
has no drift term initially, then we write the probability 
distribution function in the Gaussian form assuming the limit 
of small diffusion 
 \begin{eqnarray}
 P(\Theta)\propto \exp\left[-{\Theta^2\over 2\upsilon_{22}}\right],
\end{eqnarray}
where $\upsilon_{22}$ is given by (\ref{vtwo}).

%%%%%%%%%%%%%%%%%%%%%%%%%%%%%%%%%%%%%%%%%%%%%%%%%%%%%%%%%%%%%%%%%%%%
\section{Expansion of the probability function }
%%%%%%%%%%%%%%%%%%%%%%%%%%%%%%%%%%%%%%%%%%%%%%%%%%%%%%%%%%%%%%%%%%%%
In this Appendix, we present the expression of the probability 
function expanded in terms of $\Delta r$ and $\Delta s$. 
The variable $\xi$ is written in terms 
of $r$, equation (\ref{trr}). Then the part of the 
probability function written by $\xi$ provides us with 
information about the adiabatic component, which
can be expanded as follows, by setting $r=r_{cl}+\Delta r$, 
\begin{eqnarray}
  \exp\left[-{(\xi-\xi_*)^2\over 2\upsilon _{11}}\right]
  &=&\exp\Biggl[-R_n
  \Biggl\{\left({\Delta r\over r_{cl}}\right)^2
\nonumber\\
&& -\left(n-1+{n+2\over2}\left({r_{cl}\over r_*}\right)^{2-n}\right)
  \left({\Delta r\over r_{cl}}\right)^3 
\nonumber \\
&& +\Biggl( 
  \frac{(n-1)(7n-3)}{12}+\frac{3(n-1)(n+2)}{4}
  \left(\frac{r_{cl}}{r_*}\right)^{2-n}
\nonumber\\
&&+\frac{(n+2)(30-n)}{48}\left(\frac{r_{cl}}{r_*}
\right)^{2(2-n)} \Biggr)  \left(\frac{\Delta r}{r_{cl}}\right)^4
  \cdots\Biggr\}\Biggr],
\label{expxi}
\end{eqnarray}
for $n\ge2$, where we defined
\begin{eqnarray}
R_n={96\pi^2\over \tlambda}
  n^2  \left[ \left( \frac{r_*}{\sigma} \right)^4 -
  \left( \frac{r_{cl}}{\sigma } \right)^4 \right] ^{-1}
  \left( \frac{r_{cl}}{\sigma } \right)^{2(2-n)}.
\label{Rn}
\end{eqnarray}
On the other hand, the part of the probability function 
of $\Theta$ is expanded as 
%\begin{eqnarray}
% \exp\left[-{\Theta^2\over 2\upsilon _{22}}\right]
%  &=&\exp\Biggl[-S_n 
%\Biggl\{\left({\Delta s\over r_{cl}}\right)^2+(2n-1){\Gamma_n}
%  \left({\Delta s\over r_{cl}}\right)^3
%\nonumber\\
%&&-2(2n-1)\left({\Delta r\over r_{cl}}\right)
%  \left({\Delta s\over r_{cl}}\right)^2
%\nonumber \\
%&&+\frac{(2n-1)^2}{4}{\Gamma_n }^2\left({\Delta s\over r_{cl}}\right)^4
%\nonumber\\
%&&-(2n-1)^2{\Gamma_n }\left(\frac{\Delta r}{r_{cl}}\right) 
%\left({\Delta s\over r_{cl}}\right)^3
%\nonumber\\
%&&+(2n-1)^2\left({\Delta r\over r_{cl}}\right)^2
%  \left({\Delta s\over r_{cl}}\right)^2
%  \cdots\Biggr\}\Biggr]
%\label{exptheta}
%\end{eqnarray}
\begin{eqnarray}
 \exp\left[-{\Theta^2\over 2\upsilon _{22}}\right]
  &=&\exp\Biggl[-S_n 
  \Biggl\{\left({\Delta s\over r_{cl}}\right)^2
  +(2n-1){\Gamma_n }
  \left({\Delta s\over r_{cl}}\right)^3
\nonumber\\
  &&\hspace{2cm}-2(2n-1)\left({\Delta r\over r_{cl}}\right)
  \left({\Delta s\over r_{cl}}\right)^2 
\nonumber\\
  &&\hspace{2cm}+\frac{2n-1}{12}((14n-3){\Gamma_n }^2+8n)\left({\Delta s\over r_{cl}}\right)^4
\nonumber\\
  &&\hspace{2cm}-(2n-1)(4n-1){\Gamma_n }\left(\frac{\Delta r}{r_
{cl}}\right) \left({\Delta s\over r_{cl}}\right)^3
\nonumber\\
  &&\hspace{2cm}+(2n-1)(4n-1)\left({\Delta r\over r_{cl}}\right)^2\left({\Delta s\over r_{cl}}\right)^2
  \cdots\Biggr\}\Biggr],
\nonumber\\
\label{exptheta}
\end{eqnarray}
where we defined
\begin{eqnarray}
S_n={96\pi^2 \over \tlambda}\times\left\{
\begin{array}{ll}{\displaystyle
\left[ -\ln \left(\frac{r_{cl}}{r_*}\right)\right]^{-1} \left(\frac{r_{cl}}{\sigma }\right)^{4(1-n)} } \hspace{3cm} (n=2) &
\\ 
{\displaystyle
\frac{(n-2)n^2}{2}\left[\left(\frac{r_{cl}}{\sigma }\right)^{2(2-n)}-\left(\frac{r_*}{\sigma }\right)^{2(2-n)}\right]^{-1}
\left(\frac{r_{cl}}{\sigma }
\right)^{4(1-n)} }&\\
\hspace{7.7cm}  (n>2) &
\end{array} \right. 
\end{eqnarray}
and 
\begin{eqnarray}
{\Gamma_n }\equiv \tan\theta-\cot\theta=(1/L)^{1/2(n-1)}-L^{1/2(n-1)}.
\label{gamman}
\end{eqnarray}

%\newpage

\end{document}